\documentclass[%
 reprint,
 amsmath,amssymb,
  pra,
]{revtex4-1}

\usepackage{graphicx}
\usepackage{dcolumn}
\usepackage{bm}

\begin{document}


\title{Transfer of Spatial Reference Frame Using Singlet States and Classical Communication}

\author{Thomas B. Bahder}
\affiliation{%
Aviation and Missile Research, 
      Development, and Engineering Center, \\    
US Army RDECOM, 
Redstone Arsenal, AL 35898, 
U.S.A. \\
}%

\date{\today}

\begin{abstract}
A simple protocol is described for transferring spatial direction from Alice to Bob (two spatially separated observers) up to inversion.  The two observers are assumed to share quantum singlet states and classical communication.    The protocol assumes that Alice and Bob have complete free will (measurement independence) and is based on maximizing the Shannon mutual information between Alice and Bob's measurement outcomes.   Repeated use of this protocol for each  spatial axis of Alice allows transfer of a complete 3-dimensional reference frame, up to inversion of each of the axes.  The technological complexity of this protocol is similar to that needed for BB84 quantum key distribution, and hence is much simpler to implement than recently proposed schemes for transmission of reference frames.   A second protocol  based on a Bayesian formalism is also presented.   
\end{abstract}

\maketitle


\section{\label{Intro}Introduction}
Many technological applications require establishing a local reference frame that is spatially aligned with some predefined global reference frame.  An example is a spacecraft that must reach a target position that is specified in a predefined global reference frame.  The spacecraft must be able to align its internal reference frame with respect to an external global frame.   Historically, mechanical gyroscopes  were used, which maintained their spatial orientation with respect to the global frame.  More recently, the classical optical Sagnac effect~\mbox{\cite{Sagnac1913a,Sagnac1913b,Sagnac1914,Post1967}}, which measures {\it rotation rate} along an axis,  is being exploited in all modern rotation sensors~\cite{Lefevre1993} and their applications to inertial navigation systems~\cite{Titterton2004}.  Even more recently, much effort has been expended on experiments with quantum Sagnac interferometers, using single-photons~\cite{Bertocchi2006}, using cold atoms~\cite{Gustavson2000,Gilowski2009} and using Bose-Einstein condensates(BEC)~\cite{Gupta2005,Wang2005,Tolstikhin2005}, in efforts to improve the sensitivity to rotation of the classical optical Sagnac effect, and schemes have also been proposed to improve the sensitivity of rotation sensing using multi-photon correlations~\cite{Kolkiran2007} and using entangled particles, which are expected to have Heisenberg limited precision that scales as $1/N$, where $N$ is the number of particles~\cite{Cooper2010}.  Limitations of classical gyroscopes have been discussed in Ref~\cite{Lefevre1993} and limits of classical Sagnac effects has been discussed in terms of Shannon mutual information in Ref~\cite{Bahder2011a}. 

In a different thread, transfer of spatial orientation and alignment of reference frames has been of recent interest from the point of view of quantum information~\cite{PhysRevA.62.040101,PhysRevLett.74.1259,PhysRevLett.86.4160}. Peres and Scudo have considered quantum particles carrying angular momentum to play the role of gyroscopes and exploited such particles to transfer direction, or orientation, in space~\cite{PhysRevLett.86.4160}.  Early it was realized that spatial direction, or reference frame orientation, is a special type of quantum information named ``unspeakable quantum information"~\cite{peres2002}, which is information that cannot be transmitted by sending classical bits of information.  Instead, it requires transfer of physical particles, see Ref~\cite{RevModPhys.79.555} for a review. 

Subsequently, it was realized that a single quantum system, such as a hydrogen atom, can transmit all three axes of a Cartesian coordinate frame~\cite{PhysRevLett.87.167901,PhysRevA.68.042308}.  In such schemes,  the rotation matrix between the predefined reference frame and the local frame is determined by POVM measurements~\cite{Helstrom1976} on the exchanged quantum system.  Such schemes are essentially multi-parameter estimation methods, where the parameters are the rotation angles that will align the frames.  Such POVM measurements require implementing complicated quantum states and POVM measurements that would require multiple sensors to make simultaneous measurements, which essentially are measurements of multi-correlation functions.   In practice, such schemes are very complex to carry out experimentally in a laboratory~\cite{PhysRevLett.87.167901,peres2002,PhysRevLett.93.180503,PhysRevLett.98.120501,PhysRevA.78.052333,RevModPhys.79.555}.  

In this Section~\ref{Protocol} of this  manuscript, I describe a simple protocol to transfer a spatial reference frame from one observer to another, from Alice to Bob.   This protocol makes use of spin $s=1/2$ singlet states and hence requires nothing more than a single Stern-Gerlach apparatus as a spin analyzer, and therefore, can be easily implemented in a  laboratory with today's technology.  Specifically, this protocol does not require complicated POVM, and hence does not require measurement of multi-particle $(N>2)$ correlation functions. 

This protocol assumes that Alice and Bob have complete free will (measurement independence)~\cite{PhysRevA.73.022104,PhysRevLett.105.250404,PhysRevA.82.062117,banik2012,PhysRevA.87.062121,PhysRevA.88.032110,PhysRevLett.109.160404} to choose directions for their Stern-Gerlach spin analyzers.  This protocol can be used to transfer a single spatial direction from Alice to Bob, up to an inversion.   This protocol can be used repeatedly to transfer the orientation of three axes in order to define a local reference frame, up to inversion of each of the axes.  

In Section~\ref{Bayesian Approach},  I describe a Bayesian protocol to transfer a reference frame from Alice to Bob.   Both,  protocols  described below have the advantage that they are significantly easier to implement with today's technology than those in previous works.  The technological complexity of these protocols is similar to that needed for the BB84 quantum key distribution~\cite{BB84-1,BB84-2}, and hence is much simpler to implement than recently proposed schemes for transmission of reference frames~\cite{PhysRevLett.93.180503}. Also, the protocols presented have  some similarity to, but are different than, secret communication of a reference frame~\cite{PhysRevLett.98.120501}.  In this work, I do not investigate security against eavesdroppers. Also, Heisenberg limited resolution is not the main concern here.  Instead, technological simplicity of implementation is the main point.  

\section{\label{Protocol} Protocol Based on Shannon Mutual Information}
Consider two observers, Alice and Bob, where Alice has a predetermined spatial direction that she wants to communicate to Bob. Assume that Alice and Bob share pairs of spin $s=1/2$ particles, where each pair is entangled in a spin singlet state
\begin{equation}
|\psi_o \rangle  = \frac{1}{{\sqrt 2 }}\left( \, {| -,{\bf{n}} {\rangle _A} \,\,  | +,{\bf{n}} {\rangle _B} - | +,{\bf{n}} {\rangle _A} \,\, | -,{\bf{n}} {\rangle _B}} \, \right)
\label{singletstate}
\end{equation}
where the basis states, $| -,{\bf{n}} \rangle _A$,  $| +,{\bf{n}} \rangle _A$,  are  eigenstates of Alice's operator, ${\bf \sigma }_A \cdot {\bf{n}}_A $, and 
$| -,{\bf{n}} \rangle _B$,  $| +,{\bf{n}} \rangle _B$  are eigenstates of Bob's operator, ${\bf \sigma }_B \cdot {\bf{n}}_B $. Here, ${\bf n}_A$ and ${\bf n}_B$ are 3-dimensional unit vectors   specifying  the direction of measurement setting in a Stern-Gerlach type measurement, for Alice and Bob, respectively.   For both Alice and Bob, the operators and  states satisfy the eigenvalues equations:
\begin{equation}
{\bf{\sigma }} \cdot {\bf{n}}| \pm ,{\bf{n}}\rangle  =  \pm | \pm,{\bf{n}} \rangle 
\label{eigenstates}
\end{equation}
where ${\bf \sigma }= (\sigma_x,\sigma_y,\sigma_z)$ is the vector of Pauli matrices, 
\begin{equation}
{\sigma _x} = \left( {\begin{array}{*{20}{c}}
   0 & 1  \\
   1 & 0  \\
\end{array}} \right),\;{\sigma _y} = \left( {\begin{array}{*{20}{c}}
   0 & { - i}  \\
   i & 0  \\
\end{array}} \right),\;{\sigma _z} = \left( {\begin{array}{*{20}{c}}
   1 & 0  \\
   0 & 1  \\
\end{array}} \right) 
\label{PauliMatrices}
\end{equation}

On any bipartite quantum state, $|\psi\rangle $, Alice and Bob can each make Stern-Gerlach projective measurements.  The conditional probability, $p(a,b|{\bf x}, {\bf y})$, that Alice obtains measurement outcome $a \in \{  - 1, + 1\} $ and Bob obtains measurement outcome $b \in \{  - 1, + 1\} $, given  Alice and Bob's measurement settings were ${\bf x}$ and ${\bf y}$, respectively, is computed using the Born rule from the expectation value of the POVM operator~\cite{barnettBook2009}, ${{\hat M}_{a|{\bf{x}}}} \otimes {{\hat M}_{b|{\bf{y}}}}$, as: 
\begin{equation}
p(a,b|{\bf{x}},{\bf{y}}) = \langle {\psi}|{{\hat M}_{a|{\bf{x}}}} \otimes {{\hat M}_{b|{\bf{y}}}}|{\psi}\rangle 
\label{ConditionalProbability}
\end{equation}
where ${{\hat M}_{a|{\bf{x}}}} = |a,{\bf{x}}\rangle \langle a,{\bf{x}}|$ and ${{\hat M}_{b|{\bf{x}}}} = |b,{\bf{x}}\rangle \langle b,{\bf{x}}|$ are projective operators. The conditional probability, $p(a,b|{\bf{x}},{\bf{y}})$, should properly be expressed as  $p(a,b|{\bf{x}},{\bf{y}}, \psi)$,  conditional on the state  $|\psi \rangle$, however, I suppress the dependence in the notation.

For the singlet state $|\psi_o \rangle$ in Eq.~(\ref{singletstate}),  Alice and Bob's measurement outcomes, $a$ and $b$, are statistically correlated. 
The correlations are expressed by the conditional probability
\begin{equation}
p(a,b|{\bf{x}},{\bf{y}}) = \frac{1}{4}\left( {1 - a{\kern 1pt} b{\kern 1pt} {\kern 1pt} {\bf{x}} \cdot {\bf{y}}} \right)
\label{singletCorrelations}
\end{equation}
Alternatively, the correlations between Alice and Bob's measurement outcomes can be expressed in terms of the Shannon mutual information~\cite{Cover2006} between Alice and Bob's measurement outcomes:  
\begin{equation}
I(A;B) = \sum\limits_a {\sum\limits_b \,\, {p(a,b) \,{{\log }_2}\left[ {\frac{{p(a,b)}}{{p(a)p(b)}}} \right]} } 
\label{ShannonMutualInfo}
\end{equation}
where the marginal probability for measurement outcomes, $p(a,b)$, is obtained from integrating the joint probability, $p(a,b, {\bf x}, {\bf y})$, given by
\begin{equation}
p(a,b, {\bf x}, {\bf y}) = p(a,b | {\bf x}, {\bf y})\,\, p({\bf x}, {\bf y})
\label{JointProbability}
\end{equation}
where $p({\bf{x}},{\bf{y}})$ is the prior probability that Alice and Bob have set their analyzer directions to $\bf{x}$ and $\bf{y}$, respectively.  
The marginal probability for measurement outcomes is then given by
\begin{equation}
p(a,b) = \int {d{\Omega _x}} \int {d{\Omega _y}} \,\, p(a,b|{\bf{x}},{\bf{y}}){\kern 1pt} p({\bf{x}},{\bf{y}})
\label{ABProbability}
\end{equation}
where the prior probability is normalized
\begin{equation}
\int {d{\Omega _x}} \int {d{\Omega _y}} {\kern 1pt} {\kern 1pt} p({\bf{x}},{\bf{y}}) = 1
\label{PriorNormalization}
\end{equation}
and the integrals are over solid angles associated with unit vectors  $\bf{x}$ and $\bf{y}$, respectively.

The marginal probabilities for Alice and Bob's measurement outcomes, $p(a)$ and $p(b)$, are given by summation over the marginal probability $p(a,b)$,
\begin{equation}
\begin{array}{l}
 p(a) = \sum\limits_b {p(a,b)}  \\ 
 p(b) = \sum\limits_a {p(a,b)}  \\ 
 \end{array}
\label{marginalProbabilities}
\end{equation}
For the singlet correlations in Eq.~(\ref{singletCorrelations}), \mbox{$p(a)=p(b)=1/2$} for any arbitrary normalized distribution of measurement settings $p(x,y)$. 

\begin{figure}[t]   
\includegraphics[width=3.0in]{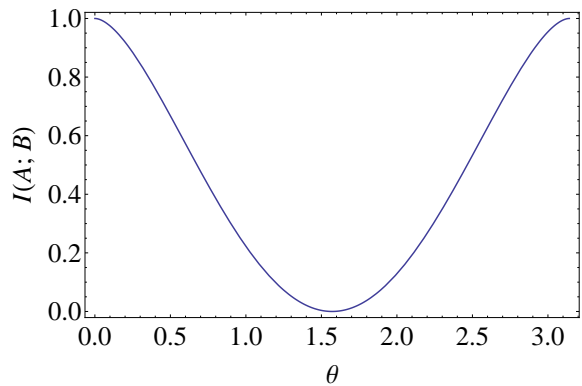}
\caption{\label{fig:SingletStateMutualInfo}The Shannon mutual information between Alice and Bob's measurement outcomes, $a$ and $b$, is plotted as a function of the angle $\theta$ between their chosen measurement directions.}
\end{figure}

I assume that Alice and Bob  have complete free will or measurement independence~\cite{PhysRevA.73.022104,PhysRevLett.105.250404,PhysRevA.82.062117,banik2012,PhysRevA.87.062121,PhysRevA.88.032110,PhysRevLett.109.160404} to choose their measurement directions.  Specifically, I assume that Alice and Bob's choice of analyzer directions $\bf{x}$ and $\bf{y}$ is not correlated, therefore, I  take the prior distribution p({\bf{x}},{\bf{y}}) as a product distribution
\begin{equation}
p({\bf{x}},{\bf{y}}) = p ({\bf{x}})  \, p({\bf{y}}) 
\label{PriorProbability}
\end{equation}
Furthermore, I assume that Alice and Bob {\it each} have unrestricted measurement freedom to choose specific directions,  ${{\bf{x}}_0}$ and ${{\bf{y}}_0}$, respectively. Therefore, I take the prior distribution to be given by Dirac $\delta$-functions:
\begin{equation}
p({\bf{x}},{\bf{y}}) = \delta ({\bf{x}} - {{\bf{x}}_o}){\kern 1pt} \delta (y - {{\bf{y}}_o})
\label{FreeProbabilities}
\end{equation}
With these choices, the Shannon mutual information becomes
\begin{equation}
I(A;B) = \frac{1}{2}{\log _2}\left[ {1 - {{\left( {{{\bf{x}}_o} \cdot {{\bf{y}}_o}} \right)}^2}} \right] - \frac{1}{2}{{\bf{x}}_o} \cdot {{\bf{y}}_o}{\log _2}\left[ {\frac{{1 - {{\bf{x}}_o} \cdot {{\bf{y}}_o}}}{{1 + {{\bf{x}}_o} \cdot {{\bf{y}}_o}}}} \right]
\label{ShannonSinglet}
\end{equation}
Since space is isotropic, the mutual information only depends on the relative angle $\theta$ between Alice and Bob's measurement settings,
\begin{equation}
\cos \theta  = {{\bf{x}}_o} \cdot {{\bf{y}}_o}
\label{relativeAngle}
\end{equation}
The mutual information in Eq.~(\ref{ShannonSinglet}) is a functional of the cosine of the angle between the vectors, 
and can be notated as $I\left[ {{{\bf{x}}_o} \cdot {{\bf{y}}_o}} \right]$.
Using the polar coordinate representation for Alice and Bob's unit vectors, in Cartesian vector components
\begin{eqnarray}
{\bf{x}}_o & = & (\sin {\theta _x}\cos {\phi _x}, \sin {\theta _x}\sin {\phi _x},   \cos {\theta _x})  \nonumber \\
{\bf{y}}_o & = & (\sin {\theta _y}\cos {\phi _y}, \sin {\theta _y}\sin {\phi _y},   \cos {\theta _y}) 
\label{PolarUnitVectors}
\end{eqnarray}
the cosine of the relative angle is written as 
\begin{eqnarray}
\cos \theta  & = & \sin {\theta _x}\cos {\phi _x}\sin {\theta _y}\cos {\phi _y} \nonumber \\
              &  & + \sin {\theta _x}\sin {\phi _x}\sin {\theta _y}\sin {\phi _y} + \cos {\theta _x}\cos {\theta _y}
\label{relativeAngleExpansion}
\end{eqnarray}
where $\theta _x$ and $\phi _x$ are the polar angles for Alice's unit vector, ${{\bf{x}}_o}$, and $\theta_y$ and $\phi_y$ are the polar angles for Bob's unit vector, ${{\bf{y}}_o}$.
If Alice holds her unit vector ${{\bf{x}}_o}$ constant, then Bob can determine Alice's  vector ${\bf y}_o$, or its negative $-{\bf x}_o$, by searching for a maximum in 
the mutual information $I\left[ {{{\bf{x}}_o} \cdot {{\bf{y}}_o}} \right]$ as a function of the two angles  $\theta_y$ and $\phi_y$ that describe Bob's vector ${{\bf{y}}_o}$.   
The mutual information in Eq.~(\ref{ShannonSinglet}) is plotted in Figure~\ref{fig:SingletStateMutualInfo}. 
The mutual information is a maximum when Alice and Bob's chosen  vectors, ${{{\bf{x}}_o}}$ and ${{{\bf{y}}_o}}$, are parallel or anti-parallel, i.e., when ${{{\bf{x}}_o} \cdot {{\bf{y}}_o}} = \pm 1$. 
\begin{figure}[t]   
\includegraphics[width=3.0in]{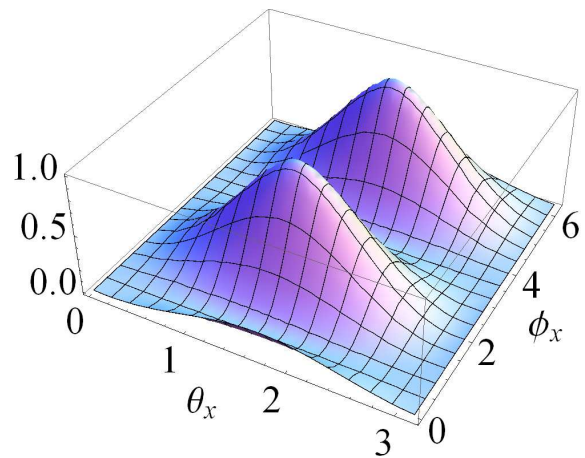}
\caption{\label{fig:MutualInfoVSAngles}The Shannon mutual information between Alice and Bob's measurement outcomes, $I\left[ {{{\bf{x}}_o} \cdot {{\bf{y}}_o}} \right]$, is plotted as a function of the polar angles $\theta_y$ and $\phi_y$ that describe Bob's unit vector, for the case where Alice's unit vector is given by $\theta _x=1.5$ and $\phi _x=2.1$. The two peaks show the degeneracy in mutual information.}
\end{figure}
This degeneracy in mutual information prevents it from being used to distinguish between these two cases.  However, in practice we often know in which hemisphere the unknown unit vector points, so this may not be a concern. Alternatively, if we have complete ignorance of the unknown unit vector, we may employ a quantum glove~\cite{gisin2004,PhysRevA.72.022304}, as mentioned below.

Assume that Alice wants to transmit to Bob the orientation of her unit vector ${\bf x}_o$.  Assume that Alice and Bob share $N$ distinguishable copies of spin singlet states, where each spin singlet state is labeled, say by integers, $1, 2, \cdots M$.   Assume that they agree to make $M$ measurements on these spin singlet states, in order of increasing integer label.  Alice chooses a fixed direction in 3-dimensional space, defined by her unit vector ${\bf x}_o$, that she wishes to transmit to Bob.  Bob chooses a trial guess unit vector, ${\bf y}_o^{(1)}$, to be the orientation of Alice's unit vector ${\bf x}_o$.   Then, Alice and Bob make the $M$ measurements on their shared singlet states.  Alice and Bob each record their measurement outcomes, which are either -1 or +1.   Alice has a string of $M$ measurement outcomes, where $M^+_A$ have value +1 and $M-M_A^+$ have value -1. By a classical channel, Alice sends to Bob  the sequence of measurement outcomes.  Similarly, Bob has a string of $M$ measurement outcomes, where $M^+_B$ have value +1 and $M-M_B^+$ have value -1.   Bob can compute an estimate of $p(a)$, defined as $\tilde p(a)$, from the data that Alice sent to him by a classical channel:
\begin{equation}
\tilde p(a =  + 1) = \frac{{M_A^ + }}{{{M}}}
\label{paPlus}
\end{equation}
Similarly Bob can estimate, $p(a=-1)$, defined as \mbox{$\tilde p(a=-1)$},
\begin{equation}
\tilde p(a =  -1) = \frac{{M_A^ - }}{{{M}}}
\label{paMinus}
\end{equation}
From his own measurement outcomes, Bob can make an estimate of $p(b=+1)$, defined as $\tilde p(b=+1)$, 
\begin{equation}
\tilde p(b =  + 1) = \frac{{M_B^ + }}{{{M}}}
\label{pbPlus}
\end{equation}
and similarly Bob can estimate, $p(b=-1)$, defined as \mbox{$\tilde p(b=-1)$},
\begin{equation}
\tilde p(b =  -1) = \frac{{M_b^ - }}{{{M}}}
\label{pbMinus}
\end{equation}
For large values of measurements, $M$, the probabilities $\tilde{p}(a)$ and $\tilde{p}(b)$ are expected to have values close to $1/2$.
Furthermore, Bob can estimate the probabilities $p(a,b)$ by comparing and counting correlations in his and Alice's measurement outcomes. From the data, Bob can compute the four numbers, 
$M^{\alpha \, \beta}_{A B}$, where $\alpha, \beta \in \{ -1,+1 \}$, where $M^{++}_{ab}$ is the number of times that both Alice and Bob had $+1$ in their measurements, $M^{-+}_{A B}$ is the number of times that Alice had $-1$ and Bob had $+1$ in their measurements, respectively, $M^{+-}_{A B}$ is the number of times that Alice had $+1$ and Bob had $-1$ in their measurements, respectively,  and $M^{--}_{A B}$ is the number of times that both Alice and Bob had $-1$ in their measurements.   Bob can then find an estimate, $\tilde p(a = \alpha ,b = \beta )$, for  the probabilities $p(a,b)$,  
\begin{equation}
\tilde p(a = \alpha, b = \beta ) = \frac{{M_{ A B}^{\alpha \beta }}}{{{M}}}
\label{pAB}
\end{equation}
where  $M=\sum_\alpha \sum_\beta {M_{A B}^{\alpha \beta }} = 1 $.

Using the estimated distributions, $\tilde p(a)$, $\tilde p(b)$, and $\tilde p(a = \alpha ,b = \beta )$, in Eq.~(\ref{ShannonMutualInfo}) in place of the actual distributions, 
$ p(a)$, $p(b)$, and $p(a,b )$, Bob can  compute an estimate of the Shannon mutual information, ${I^{est}}(A;B)$,  for the chosen pair of vectors, ${\bf x}_o$ and ${\bf y}_o$. 
In this way, Bob can determine an estimate of the mutual information, ${I^{est}}\left[ {{{\bf{x}}_o} \cdot {{\bf{y}}_o}} \right]$, as a functional of his choice of vector ${\bf y}_o$. Alice and Bob can then repeat the process $N$ times, each time Bob choosing a different direction vector ${\bf y}_o$. Bob then finds $N$ estimates for the  mutual information ${I^{est}}\left[ {{{\bf{x}}_o} \cdot {\bf{y}}_o^{(i)}} \right]$, for  $i=1, 2,\cdots, N$.  The vector ${\bf{y}}_o^{(i)}$ giving the largest mutual information is then Bob's best estimate of Alice's  vector ${\bf{x}_o}$. Of course, as mentioned above, the mutual information is a maximum when ${{\bf{x}}_o} \cdot {{\bf{y}}_o} = +1$ and ${{\bf{x}}_o} \cdot {{\bf{y}}_o} = -1$, so Bob's best estimate may by a vector that is anti-parallel to Alice's vector.  This may not be an issue in practice if Bob has sufficient initial information (up to the hemisphere) about Alice's vector.  Alternatively, if Bob is completely ignorant about Alice's vector, then he may employ a quantum glove~\cite{gisin2004,PhysRevA.72.022304} approach to determine the correct orientation from the two possibilities.  Obviously, the whole procedure above may be repeated two more times in order for Bob to determine a reference frame with three axes parallel to Alice's axes.

\section{\label{Bayesian Approach}Bayesian Approach}
The above protocol can be compared to a Bayesian approach, where a probability distribution for the unknown quantity, $ {\bf x}_o \cdot {\bf y }_o $ , is determined from the data.  In the protocol of the previous section, I assumed that Alice and Bob had the measurement freedom  to choose  directions  ${\bf x}_o$ and ${\bf y}_o$, so the prior distribution, $p({\bf x}, {\bf y})$, was given by delta functions, see Eq~(\ref{FreeProbabilities}).  In the Bayesian approach, I assume that the prior distribution is flat, $p({\bf x}, {\bf y})=1/(4 \pi)^2$, so Alice and Bob have no prior knowledge of ${\bf x}_o \cdot {\bf y}_o$.  Assume that Alice and Bob make $N$ total  measurements on singlet states, leading to the data on measurement outcomes, 
$\{ {a_i},{b_i}\}  \equiv \{ ({a_1},{b_1}),({a_2},{b_2}), \cdots ,({a_N},{b_N})\} $.    Each measurement outcome, $({a_j},{b_j})$, is independent in the sequence, so the data can be modeled by the product probability distribution
\begin{equation}
p(\{ {a_i},{b_i}\} |{\bf{x}},{\bf{y}}) = \prod\limits_{j = 1}^N {\frac{1}{4}\left( {1 - {a_j}{\kern 1pt} {b_j}{\kern 1pt} {\kern 1pt} {\bf{x}} \cdot {\bf{y}}} \right)} 
\label{ProbSequenceBayesian}
\end{equation}
Using Bayes' rule, I can write the conditional probability density for Alice and Bob's vectors, ${\bf{x}}$ and ${\bf{y}}$, given the data $\{ {a_i},{b_i}\}$ as:
\begin{equation}
p({\bf{x}},{\bf{y}}|\{ {a_i},{b_i}\} ) = \frac{{p(\{ {a_i},{b_i}\} |{\bf{x}},{\bf{y}})\,p({\bf{x}},{\bf{y}})}}{{\int {d{\Omega _x}} \int {d{\Omega _y}} {\kern 1pt} {\kern 1pt} p(\{ {a_i},{b_i}\} |{\bf{x}},{\bf{y}})\,\,p({\bf{x}},{\bf{y}})}}
\label{Prob_X.Y}
\end{equation}
Assuming complete ignorance of the angle between vectors ${\bf x}$ and ${\bf y}$ and therefore taking $p({\bf x},{\bf y})=1/(4 \pi)^2$, I find
\begin{equation}
p({\bf{x}},{\bf{y}}|\{ {a_i},{b_i}\} ) = \frac{1}{{8{\pi ^2}}}\frac{{{{\left( {1 - {\bf{x}} \cdot {\bf{y}}} \right)}^{{n_ + }}}\,{{\left( {1 + {\bf{x}} \cdot {\bf{y}}} \right)}^{{n_ - }}}}}{{d({n_ + },{n_ - })}}
\label{Prob_X.Y_1}
\end{equation}
where the function in the denominator is
\begin{equation}
\begin{array}{l}
 d({n_ + },{n_ - }) = \int\limits_{ - 1}^{ + 1} {d\gamma {\kern 1pt} } {\left( {1 - \gamma } \right)^{{n_ + }}}\,{\left( {1 + \gamma } \right)^{{n_ - }}} \\ 
  = \frac{{{}_2{F_1}(1, - {n_ - };2 + {n_ + }; - 1)}}{{1 + {n_ + }}} + \frac{{{}_2{F_1}(1, - {n_ + };2 + {n_ - }; - 1)}}{{1 + {n_ - }}} \\ 
 \end{array}
\label{d_Def}
\end{equation}
where ${}_2{F_1}(a,b;c;z)$ is the hypergeometric function~\cite{Abromowitz_Stegun} and the two functions
\begin{equation}
\begin{array}{l}
 {n_ + } = {n_ + }(({a_1},{b_1}),({a_2},{b_2}), \cdots ,({a_N},{b_N})) = \sum\limits_{k = 1}^M {{\delta _{{a_{k{\kern 1pt} }}{b_k},1}}}  \\ 
 {n_ - } = {n_ - }(({a_1},{b_1}),({a_2},{b_2}), \cdots ,({a_N},{b_N})) = \sum\limits_{k = 1}^M {{\delta _{{a_{k{\kern 1pt} }}{b_k}, - 1}}}  \\ 
 \end{array}
\label{nPlus_nMinus}
\end{equation}
count how many times  the product $ a_j b_j=+1$ and $ a_j b_j=-1$, respectively, and 
\begin{equation}
N= n_{+} + n_{-}
\label{Ndefinition}
\end{equation}

\begin{figure}[t]   
\includegraphics[width=3.0in]{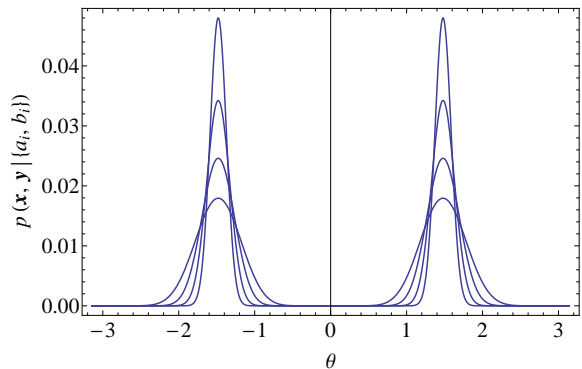}
\caption{\label{fig:Prob_xy_Plot}The probability density $p({\bf{x}},{\bf{y}}|\{ {a_i},{b_i}\} )$, given in Eq.(\ref{Prob_X.Y_3}), is plotted versus the angle $\theta$ between unit vectors  $ {\bf x}$ and $ {\bf y }$, for total number of measurements $N=11,22,44,88$.  For each $N$, there are two peaks symmetric about $\theta=0$.  The width of each peak in the plots is successively smaller with increasing number of measurements $N= n_+ + n_{-}$.  The plots have a constant ratio of $n_+ / n_{-} =5/6$.   The plot have values $(n_+ , n_{-})=\{(5,6),(10,12),(20,24),(40,48) \}$  .}
\end{figure}

The distribution in Eq.~(\ref{Prob_X.Y_1}) can also be written  in terms of the cosine of the angle between Alice and Bob's vectors
\begin{equation}
p({\bf{x}},{\bf{y}}|\{ {a_i},{b_i}\} ) = \frac{{{2^N}}}{{8{\pi ^2}}}\frac{{{{\left( {\sin \frac{\theta }{2}} \right)}^{2{n_ + }}}\,{{\left( {\cos \frac{\theta }{2}} \right)}^{2{n_ - }}}}}{{d({n_ + },{n_ - })}}
\label{Prob_X.Y_3}
\end{equation}
where $ \cos \theta = {\bf x} \cdot {\bf y } $ and $0 \le {n_ + },{n_ - } \le N$. The distribution  $p({\bf{x}},{\bf{y}}|\{ {a_i},{b_i}\} )$ satisfies the same normalization condition, given by Eq.~(\ref{PriorNormalization}), as the prior distribution $p({\bf{x}},{\bf{y}})$. Figure~\ref{fig:Prob_xy_Plot} shows a plot of this probability density.  The probability density is symmetric in angle $\theta$, so there is an ambiguity in the angle between vectors $ {\bf x}$ and $ {\bf y }$, similar to the ambiguity in the protocol based on mutual information.   As the total number of measurements $N$ increases, the peaks decrease in width. When the ratio of $n_+ / n_{-}$ changes, the peaks move farther or closer together, always remaining symmetrical about the origin. Note that the peaks are not Gaussian in shape.  One must remember that the distribution in Eq.~(\ref{Prob_X.Y_3}) is a normalized distribution when  integrated over two solid angles, see Eq.~(\ref{PriorNormalization}), and not when integrated over the angle $\theta$.

Differentiating  Eq.~(\ref{Prob_X.Y_3}) with respect to $\cos \theta$  and solving for the root shows that the peak in the distribution occurs at
\begin{equation}
\cos \theta = \frac{  n_{-} - n_{+}}{n_{-} + n_{+}}
\label{max}
\end{equation}

Using Bayes' rule, I write the probability density for Bob's unit vector from Eq.~(\ref{Prob_X.Y_3}):

\begin{eqnarray}
 p({\bf{x}}|\{ {a_i},{b_i}\} ,{\bf{y}})  & =  &  \frac{p({\bf{x}},{\bf{y}}|\{ {a_i},{b_i}\} ) }{p({\bf{y}})}  \nonumber  \\
                                         & = &     4\pi p({\bf{x}},{\bf{y}}|\{ {a_i},{b_i}\} ) 
\label{BobvectorDistribution}
\end{eqnarray}
where I have assumed that Alice's prior distribution is flat, $p({\bf{y}})=1/(4 \pi)$.  For a given data set, $\{ {a_i},{b_i}\}$,   and given vector for Alice, ${\bf{y}}$, the function $ p({\bf{x}}|\{ {a_i},{b_i}\} ,{\bf{y}})$  can be solved for two polar angles that define Bob's unit vector ${\bf{x}}$.  The distribution for Bob's vector is normalized by
\begin{equation}
\int {d{\Omega _x}} \, \, p({\bf{x}}|\{ {a_i},{b_i}\} ,{\bf{y}}) = 1
\label{BobvectorDistribution2}
\end{equation}
and ${\bf y }$, for total number of measurements 
$N=n_{+} + n_{-} = 100$, for values $(n_{+}, n_{-}) =(0,100),(10,90),(20,80)\cdots,(100,0)$. For $(n_{+},n_{-})=(0,100)$ there is one peak at $\theta=0$. 
For values of $n_{+}>0$, the one peak splits into two and moves symmetrically away from $\theta=0$ with increasing $n_{+}$. For increasing values of $n_{+}$, there are two peaks symmetric about $\theta=0$. The differing size of the peaks is due to the fact that probability density $p({\bf{x}},{\bf{y}}|\{ {a_i},{b_i}\} )$ is normalized in terms of the double integral over solid angles, given in Eq.~(\ref{PriorNormalization}), and not integration over angle $\theta$.

\section{\label{Summary}Summary}
In  previous work, it was shown that by exchanging  a single quantum system between Alice and Bob, a reference frame can be transmitted from Alice to Bob~\cite{PhysRevLett.87.167901,PhysRevA.68.042308}. This requires implementing a complex POVM.   In this paper, I show two simple protocols that can be used to transfer a reference frame (up to inversion of axes) from Alice to Bob.    These protocols require that Alice and Bob share entangled singlet states and classical communication.  The first protocol assumes that both Alice and Bob have complete free will or measurement independence~\cite{PhysRevA.73.022104,PhysRevLett.105.250404,PhysRevA.82.062117,banik2012,PhysRevA.87.062121,PhysRevA.88.032110,PhysRevLett.109.160404} to choose directions for their Stern-Gerlach spin analyzers. This protocol is based on maximizing the Shannon mutual information between Alice and Bob's measurement results.   The second protocol is based on a Bayesian approach.  In both cases, Bob can determine the spatial direction of Alice's measurement apparatus, up to a spatial inversion. If there is enough prior knowledge on Alice's measurement settings, a unique direction can be transferred.  Repeating the  protocol for each $x$, $y$ and $z$ axis of Alice allows transfer to Bob of a complete three dimensional reference frame, up to inversion of each of the axes. 
%
%


%
\bibliographystyle{apsrev}
\bibliography{References-Quantum}

\begin{thebibliography}{40}
\expandafter\ifx\csname natexlab\endcsname\relax\def\natexlab#1{#1}\fi
\expandafter\ifx\csname bibnamefont\endcsname\relax
  \def\bibnamefont#1{#1}\fi
\expandafter\ifx\csname bibfnamefont\endcsname\relax
  \def\bibfnamefont#1{#1}\fi
\expandafter\ifx\csname citenamefont\endcsname\relax
  \def\citenamefont#1{#1}\fi
\expandafter\ifx\csname url\endcsname\relax
  \def\url#1{\texttt{#1}}\fi
\expandafter\ifx\csname urlprefix\endcsname\relax\def\urlprefix{URL }\fi
\providecommand{\bibinfo}[2]{#2}
\providecommand{\eprint}[2][]{\url{#2}}

\bibitem[{\citenamefont{Sagnac}(1913{\natexlab{a}})}]{Sagnac1913a}
\bibinfo{author}{\bibfnamefont{G.}~\bibnamefont{Sagnac}},
  \bibinfo{journal}{Compt. Rend.} \textbf{\bibinfo{volume}{157}},
  \bibinfo{pages}{708} (\bibinfo{year}{1913}{\natexlab{a}}).

\bibitem[{\citenamefont{Sagnac}(1913{\natexlab{b}})}]{Sagnac1913b}
\bibinfo{author}{\bibfnamefont{G.}~\bibnamefont{Sagnac}},
  \bibinfo{journal}{Compt. Rend.} \textbf{\bibinfo{volume}{157}},
  \bibinfo{pages}{1410} (\bibinfo{year}{1913}{\natexlab{b}}).

\bibitem[{\citenamefont{Sagnac}(1914)}]{Sagnac1914}
\bibinfo{author}{\bibfnamefont{G.}~\bibnamefont{Sagnac}}, \bibinfo{journal}{J.
  Phys. Radium} \textbf{\bibinfo{volume}{5th Series 4}}, \bibinfo{pages}{177}
  (\bibinfo{year}{1914}).

\bibitem[{\citenamefont{Post}(1967)}]{Post1967}
\bibinfo{author}{\bibfnamefont{E.~J.} \bibnamefont{Post}},
  \bibinfo{journal}{Rev. Mod. Phys.} \textbf{\bibinfo{volume}{39}},
  \bibinfo{pages}{475} (\bibinfo{year}{1967}).

\bibitem[{\citenamefont{Lefevre}(1993)}]{Lefevre1993}
\bibinfo{author}{\bibfnamefont{H.}~\bibnamefont{Lefevre}},
  \emph{\bibinfo{title}{The fiber-optic gyroscope}} (\bibinfo{publisher}{Artech
  House}, \bibinfo{address}{Boston, USA}, \bibinfo{year}{1993}).

\bibitem[{\citenamefont{Titterton and Weston}(2004)}]{Titterton2004}
\bibinfo{author}{\bibfnamefont{D.~H.} \bibnamefont{Titterton}}
  \bibnamefont{and} \bibinfo{author}{\bibfnamefont{J.}~\bibnamefont{Weston}},
  \emph{\bibinfo{title}{Strapdown Inertial Navigation Technology}}
  (\bibinfo{publisher}{The Insitution of Engineering and Technology and The
  American Institute of Aeronautics}, \bibinfo{address}{London, U.K. and
  Reston, Virginia, USA}, \bibinfo{year}{2004}), \bibinfo{edition}{second
  edition} ed.

\bibitem[{\citenamefont{Bertocchi et~al.}(2006)\citenamefont{Bertocchi,
  Alibart, Ostrowsky, Tanzilli, and Baldi}}]{Bertocchi2006}
\bibinfo{author}{\bibfnamefont{G.}~\bibnamefont{Bertocchi}},
  \bibinfo{author}{\bibfnamefont{O.}~\bibnamefont{Alibart}},
  \bibinfo{author}{\bibfnamefont{D.~B.} \bibnamefont{Ostrowsky}},
  \bibinfo{author}{\bibfnamefont{S.}~\bibnamefont{Tanzilli}}, \bibnamefont{and}
  \bibinfo{author}{\bibfnamefont{P.}~\bibnamefont{Baldi}}, \bibinfo{journal}{J.
  Phys. B} \textbf{\bibinfo{volume}{39}}, \bibinfo{pages}{1011}
  (\bibinfo{year}{2006}).

\bibitem[{\citenamefont{Gustavson et~al.}(2000)\citenamefont{Gustavson,
  Landragin, and Kasevich}}]{Gustavson2000}
\bibinfo{author}{\bibfnamefont{T.~L.} \bibnamefont{Gustavson}},
  \bibinfo{author}{\bibfnamefont{A.}~\bibnamefont{Landragin}},
  \bibnamefont{and} \bibinfo{author}{\bibfnamefont{M.~A.}
  \bibnamefont{Kasevich}}, \bibinfo{journal}{Class. Quantum Grav.}
  \textbf{\bibinfo{volume}{17}}, \bibinfo{pages}{2385} (\bibinfo{year}{2000}).

\bibitem[{\citenamefont{Gilowski et~al.}(2009)\citenamefont{Gilowski, Schubert,
  Wendrich, Berg, Tackmann, Ertmer, and Rasel}}]{Gilowski2009}
\bibinfo{author}{\bibfnamefont{M.}~\bibnamefont{Gilowski}},
  \bibinfo{author}{\bibfnamefont{C.}~\bibnamefont{Schubert}},
  \bibinfo{author}{\bibfnamefont{T.}~\bibnamefont{Wendrich}},
  \bibinfo{author}{\bibfnamefont{P.}~\bibnamefont{Berg}},
  \bibinfo{author}{\bibfnamefont{G.}~\bibnamefont{Tackmann}},
  \bibinfo{author}{\bibfnamefont{W.}~\bibnamefont{Ertmer}}, \bibnamefont{and}
  \bibinfo{author}{\bibfnamefont{E.~M.} \bibnamefont{Rasel}},
  \bibinfo{journal}{Frequency Control Symposium, 2009 Joint with the 22nd
  European Frequency and Time forum. IEEE International} pp.
  \bibinfo{pages}{1173 -- 1175} (\bibinfo{year}{2009}).

\bibitem[{\citenamefont{Gupta et~al.}(2005)\citenamefont{Gupta, Murch, Moore,
  Purdy, and Stamper-Kurn}}]{Gupta2005}
\bibinfo{author}{\bibfnamefont{S.}~\bibnamefont{Gupta}},
  \bibinfo{author}{\bibfnamefont{K.~W.} \bibnamefont{Murch}},
  \bibinfo{author}{\bibfnamefont{K.~L.} \bibnamefont{Moore}},
  \bibinfo{author}{\bibfnamefont{T.~P.} \bibnamefont{Purdy}}, \bibnamefont{and}
  \bibinfo{author}{\bibfnamefont{D.~M.} \bibnamefont{Stamper-Kurn}},
  \bibinfo{journal}{Phys. Rev. Lett.} \textbf{\bibinfo{volume}{95}},
  \bibinfo{pages}{143201} (\bibinfo{year}{2005}).

\bibitem[{\citenamefont{Wang et~al.}(2005)\citenamefont{Wang, Anderson, Bright,
  Cornell, Diot, Kishimoto, Prentiss, Saravanan, Segal, and Wu}}]{Wang2005}
\bibinfo{author}{\bibfnamefont{Y.-J.} \bibnamefont{Wang}},
  \bibinfo{author}{\bibfnamefont{D.~Z.} \bibnamefont{Anderson}},
  \bibinfo{author}{\bibfnamefont{V.~M.} \bibnamefont{Bright}},
  \bibinfo{author}{\bibfnamefont{E.~A.} \bibnamefont{Cornell}},
  \bibinfo{author}{\bibfnamefont{Q.}~\bibnamefont{Diot}},
  \bibinfo{author}{\bibfnamefont{T.}~\bibnamefont{Kishimoto}},
  \bibinfo{author}{\bibfnamefont{M.}~\bibnamefont{Prentiss}},
  \bibinfo{author}{\bibfnamefont{R.~A.} \bibnamefont{Saravanan}},
  \bibinfo{author}{\bibfnamefont{S.~R.} \bibnamefont{Segal}}, \bibnamefont{and}
  \bibinfo{author}{\bibfnamefont{S.}~\bibnamefont{Wu}}, \bibinfo{journal}{Phys.
  Rev. Lett.} \textbf{\bibinfo{volume}{94}}, \bibinfo{pages}{090405}
  (\bibinfo{year}{2005}).

\bibitem[{\citenamefont{Tolstikhin et~al.}(2005)\citenamefont{Tolstikhin,
  Morishita, and Watanabe}}]{Tolstikhin2005}
\bibinfo{author}{\bibfnamefont{O.~I.} \bibnamefont{Tolstikhin}},
  \bibinfo{author}{\bibfnamefont{T.}~\bibnamefont{Morishita}},
  \bibnamefont{and} \bibinfo{author}{\bibfnamefont{S.}~\bibnamefont{Watanabe}},
  \bibinfo{journal}{Phys. Rev. A} \textbf{\bibinfo{volume}{72}},
  \bibinfo{pages}{051603(R)} (\bibinfo{year}{2005}).

\bibitem[{\citenamefont{Kolkiran and Agarwal}(2007)}]{Kolkiran2007}
\bibinfo{author}{\bibfnamefont{A.}~\bibnamefont{Kolkiran}} \bibnamefont{and}
  \bibinfo{author}{\bibfnamefont{G.~S.} \bibnamefont{Agarwal}},
  \bibinfo{journal}{Optics Express} \textbf{\bibinfo{volume}{15}},
  \bibinfo{pages}{6798} (\bibinfo{year}{2007}).

\bibitem[{\citenamefont{Cooper et~al.}(2010)\citenamefont{Cooper, Hallwood, and
  Dunningham}}]{Cooper2010}
\bibinfo{author}{\bibfnamefont{J.~J.} \bibnamefont{Cooper}},
  \bibinfo{author}{\bibfnamefont{D.~W.} \bibnamefont{Hallwood}},
  \bibnamefont{and} \bibinfo{author}{\bibfnamefont{J.~A.}
  \bibnamefont{Dunningham}}, \bibinfo{journal}{Phys. Rev. A}
  \textbf{\bibinfo{volume}{81}}, \bibinfo{pages}{043624}
  (\bibinfo{year}{2010}).

\bibitem[{\citenamefont{Bahder}(2011)}]{Bahder2011a}
\bibinfo{author}{\bibfnamefont{T.~B.} \bibnamefont{Bahder}}
  (\bibinfo{year}{2011}), \eprint{arXiv:1101.4634}.

\bibitem[{\citenamefont{Massar}(2000)}]{PhysRevA.62.040101}
\bibinfo{author}{\bibfnamefont{S.}~\bibnamefont{Massar}},
  \bibinfo{journal}{Phys. Rev. A} \textbf{\bibinfo{volume}{62}},
  \bibinfo{pages}{040101} (\bibinfo{year}{2000}),
  \urlprefix\url{http://link.aps.org/doi/10.1103/PhysRevA.62.040101}.

\bibitem[{\citenamefont{Massar and Popescu}(1995)}]{PhysRevLett.74.1259}
\bibinfo{author}{\bibfnamefont{S.}~\bibnamefont{Massar}} \bibnamefont{and}
  \bibinfo{author}{\bibfnamefont{S.}~\bibnamefont{Popescu}},
  \bibinfo{journal}{Phys. Rev. Lett.} \textbf{\bibinfo{volume}{74}},
  \bibinfo{pages}{1259} (\bibinfo{year}{1995}),
  \urlprefix\url{http://link.aps.org/doi/10.1103/PhysRevLett.74.1259}.

\bibitem[{\citenamefont{Peres and
  Scudo}(2001{\natexlab{a}})}]{PhysRevLett.86.4160}
\bibinfo{author}{\bibfnamefont{A.}~\bibnamefont{Peres}} \bibnamefont{and}
  \bibinfo{author}{\bibfnamefont{P.~F.} \bibnamefont{Scudo}},
  \bibinfo{journal}{Phys. Rev. Lett.} \textbf{\bibinfo{volume}{86}},
  \bibinfo{pages}{4160} (\bibinfo{year}{2001}{\natexlab{a}}),
  \urlprefix\url{http://link.aps.org/doi/10.1103/PhysRevLett.86.4160}.

\bibitem[{\citenamefont{Peres and Scudo}(2002)}]{peres2002}
\bibinfo{author}{\bibfnamefont{A.}~\bibnamefont{Peres}} \bibnamefont{and}
  \bibinfo{author}{\bibfnamefont{P.~F.} \bibnamefont{Scudo}}
  (\bibinfo{year}{2002}), \eprint{arXiv:quant-ph/0201017}.

\bibitem[{\citenamefont{Bartlett et~al.}(2007)\citenamefont{Bartlett, Rudolph,
  and Spekkens}}]{RevModPhys.79.555}
\bibinfo{author}{\bibfnamefont{S.~D.} \bibnamefont{Bartlett}},
  \bibinfo{author}{\bibfnamefont{T.}~\bibnamefont{Rudolph}}, \bibnamefont{and}
  \bibinfo{author}{\bibfnamefont{R.~W.} \bibnamefont{Spekkens}},
  \bibinfo{journal}{Rev. Mod. Phys.} \textbf{\bibinfo{volume}{79}},
  \bibinfo{pages}{555} (\bibinfo{year}{2007}),
  \urlprefix\url{http://link.aps.org/doi/10.1103/RevModPhys.79.555}.

\bibitem[{\citenamefont{Peres and
  Scudo}(2001{\natexlab{b}})}]{PhysRevLett.87.167901}
\bibinfo{author}{\bibfnamefont{A.}~\bibnamefont{Peres}} \bibnamefont{and}
  \bibinfo{author}{\bibfnamefont{P.~F.} \bibnamefont{Scudo}},
  \bibinfo{journal}{Phys. Rev. Lett.} \textbf{\bibinfo{volume}{87}},
  \bibinfo{pages}{167901} (\bibinfo{year}{2001}{\natexlab{b}}),
  \urlprefix\url{http://link.aps.org/doi/10.1103/PhysRevLett.87.167901}.

\bibitem[{\citenamefont{Lindner et~al.}(2003)\citenamefont{Lindner, Peres, and
  Terno}}]{PhysRevA.68.042308}
\bibinfo{author}{\bibfnamefont{N.~H.} \bibnamefont{Lindner}},
  \bibinfo{author}{\bibfnamefont{A.}~\bibnamefont{Peres}}, \bibnamefont{and}
  \bibinfo{author}{\bibfnamefont{D.~R.} \bibnamefont{Terno}},
  \bibinfo{journal}{Phys. Rev. A} \textbf{\bibinfo{volume}{68}},
  \bibinfo{pages}{042308} (\bibinfo{year}{2003}),
  \urlprefix\url{http://link.aps.org/doi/10.1103/PhysRevA.68.042308}.

\bibitem[{\citenamefont{Helstrom}(1976)}]{Helstrom1976}
\bibinfo{author}{\bibfnamefont{C.~W.} \bibnamefont{Helstrom}},
  \emph{\bibinfo{title}{Quantum Detection and Estimation Theory}}
  (\bibinfo{publisher}{Academic Press}, \bibinfo{address}{New York},
  \bibinfo{year}{1976}).

\bibitem[{\citenamefont{Chiribella et~al.}(2004)\citenamefont{Chiribella,
  D'Ariano, Perinotti, and Sacchi}}]{PhysRevLett.93.180503}
\bibinfo{author}{\bibfnamefont{G.}~\bibnamefont{Chiribella}},
  \bibinfo{author}{\bibfnamefont{G.~M.} \bibnamefont{D'Ariano}},
  \bibinfo{author}{\bibfnamefont{P.}~\bibnamefont{Perinotti}},
  \bibnamefont{and} \bibinfo{author}{\bibfnamefont{M.~F.}
  \bibnamefont{Sacchi}}, \bibinfo{journal}{Phys. Rev. Lett.}
  \textbf{\bibinfo{volume}{93}}, \bibinfo{pages}{180503}
  (\bibinfo{year}{2004}),
  \urlprefix\url{http://link.aps.org/doi/10.1103/PhysRevLett.93.180503}.

\bibitem[{\citenamefont{Chiribella et~al.}(2007)\citenamefont{Chiribella,
  Maccone, and Perinotti}}]{PhysRevLett.98.120501}
\bibinfo{author}{\bibfnamefont{G.}~\bibnamefont{Chiribella}},
  \bibinfo{author}{\bibfnamefont{L.}~\bibnamefont{Maccone}}, \bibnamefont{and}
  \bibinfo{author}{\bibfnamefont{P.}~\bibnamefont{Perinotti}},
  \bibinfo{journal}{Phys. Rev. Lett.} \textbf{\bibinfo{volume}{98}},
  \bibinfo{pages}{120501} (\bibinfo{year}{2007}),
  \urlprefix\url{http://link.aps.org/doi/10.1103/PhysRevLett.98.120501}.

\bibitem[{\citenamefont{Kolenderski and
  Demkowicz-Dobrzanski}(2008)}]{PhysRevA.78.052333}
\bibinfo{author}{\bibfnamefont{P.}~\bibnamefont{Kolenderski}} \bibnamefont{and}
  \bibinfo{author}{\bibfnamefont{R.}~\bibnamefont{Demkowicz-Dobrzanski}},
  \bibinfo{journal}{Phys. Rev. A} \textbf{\bibinfo{volume}{78}},
  \bibinfo{pages}{052333} (\bibinfo{year}{2008}),
  \urlprefix\url{http://link.aps.org/doi/10.1103/PhysRevA.78.052333}.

\bibitem[{\citenamefont{Bennett and Brassard}(1984)}]{BB84-1}
\bibinfo{author}{\bibfnamefont{C.~H.} \bibnamefont{Bennett}} \bibnamefont{and}
  \bibinfo{author}{\bibfnamefont{G.}~\bibnamefont{Brassard}}, in
  \emph{\bibinfo{booktitle}{Proceedings of IEEE International Conference on
  Computers, Systems and Signal Processing, Bangalore}}
  (\bibinfo{address}{Bangalore, India}, \bibinfo{year}{1984}), pp.
  \bibinfo{pages}{175--189}.

\bibitem[{\citenamefont{Bennett and Brassard}(1985)}]{BB84-2}
\bibinfo{author}{\bibfnamefont{C.~H.} \bibnamefont{Bennett}} \bibnamefont{and}
  \bibinfo{author}{\bibfnamefont{G.}~\bibnamefont{Brassard}},
  \bibinfo{journal}{IBM Technical Disclosure Bulletin}
  \textbf{\bibinfo{volume}{28}}, \bibinfo{pages}{3153} (\bibinfo{year}{1985}).

\bibitem[{\citenamefont{Barnett}(2009)}]{barnettBook2009}
\bibinfo{author}{\bibfnamefont{S.~M.} \bibnamefont{Barnett}},
  \emph{\bibinfo{title}{Quantum Information}} (\bibinfo{publisher}{Oxford
  University Press, Inc.}, \bibinfo{address}{New York, N.Y. USA},
  \bibinfo{year}{2009}).

\bibitem[{\citenamefont{Cover and Thomas}(2006)}]{Cover2006}
\bibinfo{author}{\bibfnamefont{T.~M.} \bibnamefont{Cover}} \bibnamefont{and}
  \bibinfo{author}{\bibfnamefont{J.~A.} \bibnamefont{Thomas}},
  \emph{\bibinfo{title}{Elements of Information Theory}}
  (\bibinfo{publisher}{J. Wiley \& Sons, Inc.}, \bibinfo{address}{Hoboken, New
  Jersey}, \bibinfo{year}{2006}), \bibinfo{edition}{second edition} ed.

\bibitem[{\citenamefont{Kofler et~al.}(2006)\citenamefont{Kofler, Paterek, and
  Brukner}}]{PhysRevA.73.022104}
\bibinfo{author}{\bibfnamefont{J.}~\bibnamefont{Kofler}},
  \bibinfo{author}{\bibfnamefont{T.}~\bibnamefont{Paterek}}, \bibnamefont{and}
  \bibinfo{author}{\bibfnamefont{i.~c.~v.} \bibnamefont{Brukner}},
  \bibinfo{journal}{Phys. Rev. A} \textbf{\bibinfo{volume}{73}},
  \bibinfo{pages}{022104} (\bibinfo{year}{2006}),
  \urlprefix\url{http://link.aps.org/doi/10.1103/PhysRevA.73.022104}.

\bibitem[{\citenamefont{Hall}(2010{\natexlab{a}})}]{PhysRevLett.105.250404}
\bibinfo{author}{\bibfnamefont{M.~J.~W.} \bibnamefont{Hall}},
  \bibinfo{journal}{Phys. Rev. Lett.} \textbf{\bibinfo{volume}{105}},
  \bibinfo{pages}{250404} (\bibinfo{year}{2010}{\natexlab{a}}),
  \urlprefix\url{http://link.aps.org/doi/10.1103/PhysRevLett.105.250404}.

\bibitem[{\citenamefont{Hall}(2010{\natexlab{b}})}]{PhysRevA.82.062117}
\bibinfo{author}{\bibfnamefont{M.~J.~W.} \bibnamefont{Hall}},
  \bibinfo{journal}{Phys. Rev. A} \textbf{\bibinfo{volume}{82}},
  \bibinfo{pages}{062117} (\bibinfo{year}{2010}{\natexlab{b}}),
  \urlprefix\url{http://link.aps.org/doi/10.1103/PhysRevA.82.062117}.

\bibitem[{\citenamefont{Banik et~al.}(2012)\citenamefont{Banik, Gazi1, Das,
  Rai, and Kunkri}}]{banik2012}
\bibinfo{author}{\bibfnamefont{M.}~\bibnamefont{Banik}},
  \bibinfo{author}{\bibfnamefont{M.~R.} \bibnamefont{Gazi1}},
  \bibinfo{author}{\bibfnamefont{S.}~\bibnamefont{Das}},
  \bibinfo{author}{\bibfnamefont{A.}~\bibnamefont{Rai}}, \bibnamefont{and}
  \bibinfo{author}{\bibfnamefont{S.}~\bibnamefont{Kunkri}},
  \bibinfo{journal}{J. Phys. A: Math. Theor.} \textbf{\bibinfo{volume}{45}},
  \bibinfo{pages}{205301} (\bibinfo{year}{2012}).

\bibitem[{\citenamefont{Thinh et~al.}(2013)\citenamefont{Thinh, Sheridan, and
  Scarani}}]{PhysRevA.87.062121}
\bibinfo{author}{\bibfnamefont{L.~P.} \bibnamefont{Thinh}},
  \bibinfo{author}{\bibfnamefont{L.}~\bibnamefont{Sheridan}}, \bibnamefont{and}
  \bibinfo{author}{\bibfnamefont{V.}~\bibnamefont{Scarani}},
  \bibinfo{journal}{Phys. Rev. A} \textbf{\bibinfo{volume}{87}},
  \bibinfo{pages}{062121} (\bibinfo{year}{2013}),
  \urlprefix\url{http://link.aps.org/doi/10.1103/PhysRevA.87.062121}.

\bibitem[{\citenamefont{Pope and Kay}(2013)}]{PhysRevA.88.032110}
\bibinfo{author}{\bibfnamefont{J.~E.} \bibnamefont{Pope}} \bibnamefont{and}
  \bibinfo{author}{\bibfnamefont{A.}~\bibnamefont{Kay}},
  \bibinfo{journal}{Phys. Rev. A} \textbf{\bibinfo{volume}{88}},
  \bibinfo{pages}{032110} (\bibinfo{year}{2013}),
  \urlprefix\url{http://link.aps.org/doi/10.1103/PhysRevA.88.032110}.

\bibitem[{\citenamefont{Koh et~al.}(2012)\citenamefont{Koh, Hall, Setiawan,
  Pope, Marletto, Kay, Scarani, and Ekert}}]{PhysRevLett.109.160404}
\bibinfo{author}{\bibfnamefont{D.~E.} \bibnamefont{Koh}},
  \bibinfo{author}{\bibfnamefont{M.~J.~W.} \bibnamefont{Hall}},
  \bibinfo{author}{\bibnamefont{Setiawan}},
  \bibinfo{author}{\bibfnamefont{J.~E.} \bibnamefont{Pope}},
  \bibinfo{author}{\bibfnamefont{C.}~\bibnamefont{Marletto}},
  \bibinfo{author}{\bibfnamefont{A.}~\bibnamefont{Kay}},
  \bibinfo{author}{\bibfnamefont{V.}~\bibnamefont{Scarani}}, \bibnamefont{and}
  \bibinfo{author}{\bibfnamefont{A.}~\bibnamefont{Ekert}},
  \bibinfo{journal}{Phys. Rev. Lett.} \textbf{\bibinfo{volume}{109}},
  \bibinfo{pages}{160404} (\bibinfo{year}{2012}),
  \urlprefix\url{http://link.aps.org/doi/10.1103/PhysRevLett.109.160404}.

\bibitem[{\citenamefont{Gisin}(2004)}]{gisin2004}
\bibinfo{author}{\bibfnamefont{N.}~\bibnamefont{Gisin}} (\bibinfo{year}{2004}),
  \eprint{arXiv:quant-ph/0408095}.

\bibitem[{\citenamefont{Collins et~al.}(2005)\citenamefont{Collins, Di\'osi,
  Gisin, Massar, and Popescu}}]{PhysRevA.72.022304}
\bibinfo{author}{\bibfnamefont{D.}~\bibnamefont{Collins}},
  \bibinfo{author}{\bibfnamefont{L.}~\bibnamefont{Di\'osi}},
  \bibinfo{author}{\bibfnamefont{N.}~\bibnamefont{Gisin}},
  \bibinfo{author}{\bibfnamefont{S.}~\bibnamefont{Massar}}, \bibnamefont{and}
  \bibinfo{author}{\bibfnamefont{S.}~\bibnamefont{Popescu}},
  \bibinfo{journal}{Phys. Rev. A} \textbf{\bibinfo{volume}{72}},
  \bibinfo{pages}{022304} (\bibinfo{year}{2005}),
  \urlprefix\url{http://link.aps.org/doi/10.1103/PhysRevA.72.022304}.

\bibitem[{\citenamefont{Abromowitz and Stegun}(1972)}]{Abromowitz_Stegun}
\bibinfo{author}{\bibfnamefont{M.}~\bibnamefont{Abromowitz}} \bibnamefont{and}
  \bibinfo{author}{\bibfnamefont{I.}~\bibnamefont{Stegun}},
  \emph{\bibinfo{title}{Handbook of Mathematical Functions}}
  (\bibinfo{publisher}{Dover Publications, Inc.}, \bibinfo{address}{New York,
  N.Y. USA}, \bibinfo{year}{1972}), \bibinfo{edition}{tenth printing} ed.

\end{thebibliography}
%
\end{document}